\documentclass[preprintnumbers,amsmath,amssymbm,prd]{revtex4}
\usepackage{epsfig}
\usepackage{graphicx}

\begin{document}
\title{The Hawking paradox and the Bekenstein resolution in higher-dimensional spacetimes}
\author{Shahar Hod}
\address{The Ruppin Academic Center, Emeq Hefer 40250, Israel}
\address{ }
\address{The Hadassah Institute, Jerusalem 91010, Israel}
\date{\today}

\begin{abstract}
\ \ \ The black-hole information puzzle, first discussed by Hawking
four decades ago, has attracted much attention over the years from
both physicists and mathematicians. One of the most intriguing
suggestions to resolve the information paradox is due to Bekenstein,
who has stressed the fact that the low-energy part of the
semi-classical black-hole emission spectrum is partly blocked by the
curvature potential that surrounds the black hole. As explicitly
shown by Bekenstein, this fact implies that the grey-body emission
spectrum of a (3+1)-dimensional black hole is considerably less
entropic than the corresponding radiation spectrum of a perfectly
thermal black-body emitter. Using standard ideas from quantum
information theory, it was shown by Bekenstein that, in principle,
the filtered Hawking radiation emitted by a (3+1)-dimensional
Schwarzschild black hole may carry with it a substantial amount of
information, the information which was suspected to be lost. It is
of physical interest to test the general validity of the
``information leak" scenario suggested by Bekenstein as a possible
resolution to the Hawking information puzzle. To this end, in the
present paper we analyze the semi-classical entropy emission
properties of {\it higher}-dimensional black holes. In particular,
we provide evidence that the characteristic Hawking quanta of
$(D+1)$-dimensional Schwarzschild black holes in the large $D\gg1$
regime are almost unaffected by the spacetime curvature outside the
black-hole horizon. This fact implies that, in the large-$D$ regime,
the Hawking black-hole radiation spectra are almost purely thermal,
thus suggesting that the emitted quanta {\it cannot} carry the
amount of (non-thermal) information which is required in order to
resolve the information paradox. Our analysis therefore suggests
that the elegant information leak scenario suggested by Bekenstein,
which is based on the effective grey-body (rather than a black-body)
nature of the black-hole emission spectra, {\it cannot} provide a
generic resolution to the intriguing Hawking information paradox.
\end{abstract}
\bigskip
\maketitle

\newpage


\section{The Hawking black-hole information puzzle}

The black-hole evaporation phenomenon, first predicted by Hawking
\cite{Haw1} more than four decades ago, imposes a great challenge to
our understanding of the interplay between gravity and quantum
theory. In particular, Hawking's semi-classical analysis \cite{Haw1}
asserts that black holes which were formed from the gravitational
collapse of pure quantum states will emit thermally distributed
radiation and thus eventually evolve into mixed thermal states. This
intriguing physical scenario is in sharp contradiction with the
fundamental quantum mechanical principle of unitary temporal
evolution, which asserts that pure quantum states should always
remain pure as they evolve in time \cite{Ehr}.

The incompatibility between gravity and quantum physics, which is
realized most dramatically in the Hawking evaporation process of
black holes, may also be discussed in terms of the fundamental
principles of quantum information theory \cite{Bek1}. In particular,
it is well known that perfectly thermal (black-body) radiation
cannot convey detailed information about the physical properties of
its emitting body. Thus, according to Hawking's semi-classical
analysis \cite{Haw1}, the information hidden in the intermediate
black-hole state about the initial quantum state of the collapsed
matter is lost forever with the complete thermal evaporation of the
black hole. This physically intriguing scenario is known as the
Hawking black-hole information puzzle.

\section{The elegant resolution suggested by Bekenstein}

Several physical scenarios have been suggested in order to resolve
the Hawking black-hole information puzzle, see e.g.
\cite{Bek1,Har,Pre} for excellent reviews. In the present paper we
would like to analyze a particular intriguing resolution originally
proposed by Bekenstein \cite{Bek1}. The possible solution suggested
by Bekenstein \cite{Bek1} to the Hawking information paradox
\cite{Haw1} belongs to the family of ``information leak" scenarios.
According to this suggested resolution, the information about the
initial quantum state of the collapsed matter, which is supposed to
be lost during the semi-classical Hawking evaporation process, is
actually encoded into the emitted black-hole radiation quanta
\cite{Bek1}.

Specifically, Bekenstein \cite{Bek1} has correctly pointed out that,
due to the effective curvature potential that surrounds the emitting
black hole, the Hawking radiation spectrum of a $(3+1)$-dimensional
Schwarzschild black hole departs from the familiar purely thermal
radiation spectrum of a perfect black-body emitter. In particular,
the characteristic curvature (scattering) potential of the
black-hole spacetime [see Eq. (\ref{Eq10}) below] partly {\it
blocks} the low-frequency part of the semi-classical Hawking
emission spectrum. The departure of the black-hole radiation
spectrum from the purely thermal spectrum of a perfect black-body
emitter can be quantified by the dimensionless energy-dependent
gray-body factors $\{\Gamma(\omega)\}$ \cite{Page} of the composed
black-hole-radiation-fields system. In particular, the low-frequency
part of the semi-classical black-hole radiation spectra is known to
be characterized by the simple limiting behavior \cite{Page,Noterhn}
\begin{equation}\label{Eq1}
\Gamma(\omega r_{\text{H}})\to 0\ \ \ \ \text{for}\ \ \ \ \omega
r_{\text{H}}\to 0\  .
\end{equation}

The characteristic relation (\ref{Eq1}) reflects the physically
interesting fact that, due to the effective curvature potential of
the black-hole spacetime, the low frequency part of the Hawking
black-hole emission spectra is characterized by occupation numbers
which are smaller than the corresponding occupation numbers of a
purely thermal black-body radiation \cite{Bek1}.

In his highly interesting work \cite{Bek1}, Bekenstein stressed the
fact that, due to the partial backscattering of the emitted quanta
by the effective curvature potential that surrounds the
$(3+1)$-dimensional evaporating black hole, the Hawking black-hole
(BH) emission spectrum is {\it less} entropic than the corresponding
purely thermal emission spectrum of a perfect black-body (BB)
emitter with the same radiation power $P$. In particular,
$(3+1)$-dimensional evaporating black holes are characterized by the
relation \cite{Bek1}
\begin{equation}\label{Eq2}
S^{3D}_{\text{BH-radiation}}(P)< S^{3D}_{\text{BB-radiation}}(P)\  .
\end{equation}

Using standard ideas from quantum information theory, it was pointed
out by Bekenstein \cite{Bek1} that the entropic deficiency of the
semi-classical black-hole radiation spectrum, as described by the
characteristic inequality (\ref{Eq2}), implies that the emitted
Hawking quanta may carry with them a substantial amount of
information. In particular, as intriguingly discussed by Bekenstein
\cite{Bek1} (see also the pioneering work \cite{Leb}), the maximum
rate at which information can be recovered from the ({\it
non}-thermal) black-hole emission spectrum is given by the
difference \cite{Noteunit}
\begin{equation}\label{Eq3}
\dot I^{3D}_{\text{max}}=\dot S^{3D}_{\text{BB-radiation}}(P)-\dot
S^{3D}_{\text{BH-radiation}}(P)\
\end{equation}
between the entropy outflow rate $\dot S^{3D}_{\text{BB-radiation}}$
from a perfect black-body emitter and the corresponding entropy
outflow rate $\dot S^{3D}_{\text{BH-radiation}}$ which characterizes
the (partially filtered \cite{Notefil}) Hawking semi-classical
radiation spectrum, both with the same radiation power \cite{Bek1}.
Using the relation (\ref{Eq3}), Bekenstein \cite{Bek1} has provided
compelling evidence that, for evaporating $(3+1)$-dimensional
Schwarzschild black holes, the maximum information outflow $\dot
I^{3D}_{\text{max}}$ [as defined by (\ref{Eq3})] may actually exceed
the entropy outflow in the Hawking black-hole radiation spectrum.
That is,
\begin{equation}\label{Eq4}
\dot I^{3D}_{\text{max}}>\dot S^{3D}_{\text{BH-radiation}}\
\end{equation}
for semi-classical $(3+1)$-dimensional Schwarzschild black holes.

Recalling the general relation $\Delta I=-\Delta S$ between
information and entropy \cite{Inf1,Inf2}, Bekenstein \cite{Bek1} has
stressed the intriguing fact that the characteristic
$(3+1)$-dimensional inequality (\ref{Eq4}) implies that, given an
appropriate quantum mechanism which codes the information in the
emitted Hawking quanta, this information may reduce the uncertainty
(that is, the lack of information) about the internal quantum state
of the Hawking black-hole radiation \cite{Bek1}. Thus, the
black-hole radiation can, in principle, end up in a pure quantum
state, in accord with the known fundamental principles of quantum
physics \cite{Ehr}. This is the essence of the proposed
$(3+1)$-dimensional Bekenstein resolution to the Hawking information
puzzle \cite{Bek1}.

\section{The Bekenstein resolution: Insights from the Hawking emission spectra of higher-dimensional
black holes}

It is of considerable physical interest to test the general validity
of the intriguing $(3+1)$-dimensional conclusion reached by
Bekenstein \cite{Bek1} regarding the amount of quantum information
that, in principle, can be carried by the emitted Hawking quanta. In
particular, one naturally wonders whether the information-entropy
inequality $\dot I^{3D}_{\text{max}}>\dot
S^{3D}_{\text{BH-radiation}}$ [see (\ref{Eq4})], which characterizes
the semi-classical radiation spectrum of a $(3+1)$-dimensional
Schwarzschild black hole \cite{Bek1}, is a generic property of the
Hawking emission spectra of {\it all} $(D+1)$-dimensional
Schwarzschild black holes.

In order to address this interesting physical question, in the
present paper we shall analyze the semi-classical entropy emission
properties of {\it higher}-dimensional black holes. In particular,
below we shall provide evidence that the characteristic
semi-classical Hawking radiation spectra of $(D+1)$-dimensional
Schwarzschild black holes in the large $D\gg1$ regime are
characterized by the opposite inequality $\dot
I_{\text{max}}(D\gg1)<\dot S_{\text{BH-radiation}}(D\gg1)$. This
characteristic large-$D$ information-entropy relation implies that
the emitted Hawking quanta of higher-dimensional Schwarzschild black
holes in the large $D\gg1$ regime {\it cannot} carry the amount of
information which is required in order to solve the intriguing
Hawking paradox. Our analysis (to be presented below) therefore
suggests that the elegant ``information leak" scenario proposed by
Bekenstein \cite{Bek1} more than two decades ago {\it cannot}
provide a generic resolution \cite{Notegen} to the Hawking
information puzzle.

We first recall that the energy emission rate per one bosonic degree
of freedom out of a (D+1)-dimensional black hole is given by the
integral relation \cite{Haw1,Page,ZuKa,CKK}
\begin{equation}\label{Eq5}
P_{\text{BH}}(D)={{\hbar}\over{2^{D-1}\pi^{D/2}\Gamma(D/2)}}\sum_{j}{\int_0^{\infty}}\Gamma
{{\omega^D\ d\omega}\over{{e^{\hbar\omega/T_{\text{BH}}}-1}}}\  ,
\end{equation}
where $j$ stands for the dimensionless angular momentum indices of
the emitted Hawking quanta and $\Gamma=\Gamma(\omega;j,D)$ are the
energy-dependent grey-body factors of the composed
black-hole-radiation-fields system \cite{Page}. These dimensionless
barrier penetration factors quantify the imprint of passage of the
emitted Hawking quanta through the effective curvature potential
that surrounds the radiating black hole. The physical parameter
\begin{equation}\label{Eq6}
T_{\text{BH}}={{(D-2)\hbar}\over{4\pi r_{\text{H}}}}\
\end{equation}
in (\ref{Eq5}) is the characteristic Bekenstein-Hawking temperature
of the evaporating (D+1)-dimensional Schwarzschild black hole
\cite{Noterh,SchTang}.

As pointed out in \cite{Hodcq}, the (D+1)-dimensional thermal
distribution $\omega^{D}/(e^{\hbar\omega/T_{\text{BH}}}-1)$ [see Eq.
(\ref{Eq5})] has a sharp peak at the characteristic black-hole
emission frequency
\begin{equation}\label{Eq7}
\omega^{\text{peak}}(D)={{DT}\over{\hbar}}[1+O(e^{-D})]\  .
\end{equation}
Substituting the semi-classical Bekenstein-Hawking temperature
(\ref{Eq6}) of the evaporating $(D+1)$-dimensional black holes into
(\ref{Eq7}), one finds the strong inequality
\begin{equation}\label{Eq8}
\omega(D)\times r_{\text{H}}={{D^2}\over{4\pi}}[1+O(D^{-1})]\gg1\
\end{equation}
for the characteristic Hawking quanta emitted by the
(D+1)-dimensional Schwarzschild black holes in the large $D\gg1$
regime.

The strong inequality (\ref{Eq8}) implies that, in the large $D\gg1$
regime, the typical wavelengths in the Hawking black-hole emission
spectra are very {\it short} on the length-scale set by the
spacetime curvature \cite{Hodcq}. This physically interesting fact
suggests that, in the large-$D$ regime, the emitted Hawking quanta
are almost unaffected by the spacetime curvature outside the
black-hole horizon.

In particular, it is important to emphasize the fact that the
dynamics of the emitted Hawking fields in the curved black-hole
spacetime is governed by the Schr\"odinger-like differential
equation \cite{Haw1,Page,ZuKa,CKK}
\begin{equation}\label{Eq9}
\Big({{d^2}\over{dr^2_*}}+\omega^2-V_{\text{BH}}\Big)\phi=0\  ,
\end{equation}
where $r_*$ is the familiar `tortoise' radial coordinate
\cite{ZuKa,CKK} and, for a massless perturbation field of
dimensionless angular harmonic index $l$, the effective
$(D+1)$-dimensional black-hole curvature potential in (\ref{Eq9}) is
given by \cite{CKK,Notepp}
\begin{equation}\label{Eq10}
V(r;D)=\Big[1-\Big({{r_{\text{H}}}\over{r}}\Big)^{D-2}\Big]\Big[{{l(l+D-2)+(D-1)(D-3)/4}\over{r^2}}
+{{(1-p^2)(D-1)^2r^{D-2}_{\text{H}}}\over{4r^D}}\Big]\  .
\end{equation}
Taking cognizance of the fact that the effective curvature potential
(\ref{Eq10}) of the $(D+1)$-dimensional black-hole spacetime is
characterized by the asymptotic large $D\gg1$ behavior
\begin{equation}\label{Eq11}
V^{\text{max}}_{\text{BH}}(D\gg1)=O\Big({{D^2}\over{r^2_{\text{H}}}}\Big)\
,
\end{equation}
one concludes that the typical emitted field quanta that constitute
the Hawking black-hole radiation spectra are characterized by the
dimensionless strong inequality [see Eqs. (\ref{Eq8}) and
(\ref{Eq11})] \cite{Hodcq}
\begin{equation}\label{Eq12}
{{\omega^2}\over{V^{\text{max}}_{\text{BH}}}}=O(D^2)\gg1\ \ \ \
\text{for}\ \ \ \ D\gg1\
\end{equation}
in the large $D\gg1$ regime.

The characteristic strong inequality (\ref{Eq12}) reflects the
physically intriguing fact that, in the large $D\gg1$ regime, the
propagation of the emitted Hawking fields in the black-hole exterior
region is practically {\it unaffected} by the curvature potential
[see Eqs. (\ref{Eq9}) and (\ref{Eq12})]. In particular, this
characteristic large-$D$ behavior can be quantified by the compact
dimensionless relation
\begin{equation}\label{Eq13}
\Gamma(\omega^{\text{peak}}r_{\text{H}};j,D\to\infty)\to 1^{-}\
\end{equation}
for the barrier penetration factors (grey-body factors) of the
composed $(D+1)$-dimensional black-hole-radiation-fields system.

The large-$D$ asymptotic behavior (\ref{Eq13}) implies that
higher-dimensional evaporating black holes in the large $D\gg1$
regime are characterized by almost perfect black-body ({\it
thermal}) emission spectra. In particular, this physically
interesting fact can be quantified by the dimensionless large-$D$
relation
\begin{equation}\label{Eq14}
{{\dot S_{\text{BH-radiation}}(P;D\to\infty)}\over{\dot
S_{\text{BB-radiation}}(P;D\to\infty)}}\to 1^{-}\  ,
\end{equation}
where \cite{Hodcq}
\begin{equation}\label{Eq15}
\dot
S_{\text{BB-radiation}}(P;D\to\infty)=\Big({{8\pi}\over{e}}\Big)^{1/2}\Big({{D}\over{4\pi}}\Big)^{(D+1)/2}
\times\Big({{P}\over{\hbar}}\Big)^{1/2}\
\end{equation}
is the characteristic entropy emission rate out of a
$(D+1)$-dimensional perfect black-body (thermal) emitter with a
given radiation power $P$.

Interestingly, and most importantly for our analysis, the compact
dimensionless ratio (\ref{Eq14}) yields the large-$D$ asymptotic
relation
\begin{equation}\label{Eq16}
\dot I_{\text{max}}(D\to\infty)=\dot S_{\text{BB-radiation}}(P)-\dot
S_{\text{BH-radiation}}(P)\to 0\
\end{equation}
for the maximum rate at which information can be recovered from the
Hawking black-hole radiation spectra in the large $D\gg1$ limit. As
emphasized above (see, in particular, the discussion in Sec. II), a
{\it thermally} distributed radiation spectrum, which is
characterized by the relation (\ref{Eq16}), {\it cannot} carry with
it the missing information about the initial quantum state of the
collapsed matter fields.

\section{Summary and Discussion}

One of the most promising solutions to the Hawking information
puzzle \cite{Haw1} has been raised by Bekenstein more than two
decades ago \cite{Bek1}. In his physically intriguing work
\cite{Bek1}, Bekenstein has stressed the fact that the black-hole
emission spectrum is partly blocked by the effective curvature
potential [see Eq. (\ref{Eq10})] that surrounds the emitting black
hole. This simple fact implies that the Hawking emission spectrum of
a (3+1)-dimensional black hole is considerably less entropic than
the corresponding radiation spectrum of a perfectly thermal
black-body emitter \cite{Bek1}. Using standard ideas from quantum
information theory, it was shown by Bekenstein \cite{Bek1} that the
filtered Hawking radiation emitted by a (3+1)-dimensional
Schwarzschild black hole may, in principle, carry with it a
substantial amount of information, the information which was
suspected to be lost. In particular, a $(3+1)$-dimensional black
hole is characterized by the relation [see Eq. (\ref{Eq4})]
\cite{Bek1}
\begin{equation}\label{Eq17}
\dot I^{3D}_{\text{max}}>\dot S^{3D}_{\text{BH-radiation}}\  .
\end{equation}

One naturally wonders whether the information-entropy relation
(\ref{Eq17}), which characterizes the Hawking emission spectrum of a
$(3+1)$-dimensional Schwarzschild black hole \cite{Bek1}, is a
generic property of the radiation spectra of {\it all}
$(D+1)$-dimensional Schwarzschild black holes? In order to address
this physically interesting question, in the present paper we have
analyzed the entropy emission properties of semi-classical {\it
higher}-dimensional black holes.

In particular, we have examined the intriguing ``information leak"
resolution suggested by Bekenstein in the context of
higher-dimensional gravitational theories. Taking cognizance of the
fact that $(D+1)$-dimensional Schwarzschild black holes in the large
$D\gg1$ regime behave as almost perfect black-body emitters [see
Eqs. (\ref{Eq12}) and (\ref{Eq13})], we have stressed the
interesting fact that the Hawking radiation spectra of these
higher-dimensional black holes are characterized by the inequality
\begin{equation}\label{Eq18}
\dot I_{\text{max}}(D\gg1)<\dot S_{\text{BH-radiation}}(D\gg1)\ .
\end{equation}
The characteristic relation (\ref{Eq18}) implies, in particular,
that the emitted Hawking quanta of the $(D+1)$-dimensional
Schwarzschild black holes in the large $D\gg1$ regime cannot carry
the missing information about the initial quantum state of the
collapsed matter fields.

Our compact analysis therefore suggests that, for higher-dimensional
evaporating black holes in the large $D\gg1$ regime, the elegant
``information leak" scenario proposed by Bekenstein \cite{Bek1} more
than two decades ago {\it cannot} provide a generic \cite{Notegen}
solution to the Hawking information puzzle.

\bigskip
\noindent
{\bf ACKNOWLEDGMENTS}
\bigskip

This research is supported by the Carmel Science Foundation. I thank
Yael Oren, Arbel M. Ongo, Ayelet B. Lata, and Alona B. Tea for
stimulating discussions.



\begin{thebibliography}{99}
\bibitem{Noterhn} Here $r_{\text{H}}$ is the horizon radius of the
emitting black hole.
\bibitem{Noteunit} We use gravitational units in which $G=c=k_{\text{B}}=1$.
\bibitem{Notefil} It is worth emphasizing again that the curvature potential
barrier that surrounds the evaporating black hole partly blocks the
emitted Hawking radiation fields \cite{Bek1,Page}.
\bibitem{Noterh} Here $r_{\text{H}}$ is the black-hole horizon radius, which is given
by the $(D+1)$-dimensional relation $r_{\text{H}}={[{{16\pi
M}/{(D-1)A_{D-1}}}]}^{1/{(D-2)}}$, where $M$ is the ADM mass of the
black-hole spacetime, and $A_{D-1}={{2\pi^{D/2}}/{\Gamma(D/2)}}$ is
the generalized area of a unit $(D-1)$-dimensional sphere.
\end{thebibliography}
\end{document}